\documentstyle[prb,aps]{revtex}

\begin{document}
\draft
\title{Fracton pairing mechanism for ``strange" superconductors:
       Self-assembling organic polymers and copper-oxide compounds}
%%% \author{Alexander V. Milovanov}
%%% \address{Department of Space Plasma Physics,
%%%          Space Research Institute, 117997 Moscow, Russia}
\author{Alexander V. Milovanov$^*$ and Jens J. Rasmussen}
\address{Optics and Fluid Dynamics Department, Risoe National Laboratory,
     DK-4000 Roskilde, Denmark}
\address{$^*$ permanent address:
         Space Research Institute, Profsoyuznaya street, 84/32, 
         117997 Moscow, Russia} %%%% e.mail: amilovan@mx.iki.rssi.ru

%\date{\today}
\maketitle

\begin{abstract}
%%%%%%%%%%%%%%%%%%%%%%%%%%%%%%%%%%%%%%%%%%%%%%%%%%%%%%%%%%%%%%%%%%%%%%%%%%%%%%
%%%%%%%%%%%%%%%%%%%%%%%%%%%%%%%%%%%%%%%%%%%%%%%%%%%%%%%%%%%%%%%%%%%%%%%%%%%%%%
Self-assembling organic polymers and copper-oxide compounds 
are two classes of ``strange" superconductors, whose challenging 
behavior does not comply with the traditional picture of Bardeen, 
Cooper, and Schrieffer (BCS) superconductivity in regular crystals. 
In this paper, we propose a theoretical model that accounts for the 
strange superconducting properties of either class of the materials. 
These properties are considered as interconnected manifestations of 
the same phenomenon: We argue that superconductivity occurs in the 
both cases because the charge carriers (i.e., electrons or holes) 
exchange {\it fracton excitations}, quantum oscillations of fractal 
lattices that mimic the complex microscopic organization of the 
strange superconductors. For the copper oxides, the superconducting 
transition temperature $T_c$ as predicted by the fracton mechanism is of 
the order of $\sim 150$ K. We suggest that the marginal ingredient of the 
high-temperature superconducting phase is provided by fracton coupled 
holes that condensate in the conducting copper-oxygen planes owing to 
the intrinsic field-effect-transistor configuration of the cuprate
compounds. For the gate-induced superconducting phase in the electron-doped 
polymers, we simultaneously find a rather modest transition temperature 
of $\sim (2-3)$ K owing to the limitations imposed by the electron 
tunneling processes on a fractal geometry. We speculate that 
hole-type superconductivity observes larger onset temperatures 
when compared to its electron-type counterpart. This promises an 
intriguing possibility of the high-temperature superconducting states 
in hole-doped complex materials. The theoretical methods applied in 
our study bring together the so-called {\it strange} (or 
{\it fractional}) dynamics and the unconventional, {\it topological} 
description of the complex fractal sets underlying the fracton 
spectrum. A generalized kinetic equation containing integer time 
and fractional real-space derivatives is found for the fracton 
excitations in the harmonic approximation. The fracton 
superconductivity mechanism is further discussed in connection with 
experimental observations. An important prediction of the present 
study is universality of ac conduction of the strange materials 
above the superconducting transition temperature $T_c$.
%%%%%%%%%%%%%%%%%%%%%%%%%%%%%%%%%%%%%%%%%%%%%%%%%%%%%%%%%%%%%%%%%%%%%%%%%%%%%%
%%%%%%%%%%%%%%%%%%%%%%%%%%%%%%%%%%%%%%%%%%%%%%%%%%%%%%%%%%%%%%%%%%%%%%%%%%%%%%
\end{abstract}

\pacs{61.43.-j, 63.22.+m, 74.10.+v, 72.80.Le, 74.20.Mn}

\twocolumn
%\narrowtext

%%%%%%%%%%%%%%%%%%%%%%%%%%%%%%%%%%%%%%%%%%%%%%%%%%%%%%%%%%%%%%%%%%%%%%%%%%%%%%
%%%%%%%%%%%%%%%%%%%%%%%%%%%%%%%%%%%%%%%%%%%%%%%%%%%%%%%%%%%%%%%%%%%%%%%%%%%%%%

\section{Introduction}

The physics of two-dimensional electron gases has 
been influential in condensed-matter research since 
the discovery of highly correlated liquid states, such 
as electron liquids with fractionally charged excitations, 
\cite{Fraccharge} and high-temperature superconducting fluid 
phases in copper-oxide compounds. \cite{High} The trademark of
copper oxides is challenging superconducting transition temperature 
\cite{High} (up to $T_c\sim 160$ K); this somehow involves the
specific microscopic organization of the cuprate superconductors
composed of the alternating conducting-insulating planes. 

The fundamental properties 
of correlated electrons have received attention with the 
observation of quantum Hall effect in polymeric materials (such 
as pentracene) \cite{Hall} and superconducting behavior in 
molecular crystals \cite{Molecular} and organic polymer films. 
\cite{Schon} The discovery of superconducting organic polymers 
\cite{Schon} addressed the correlations between the current-carrying 
electrons in presence of the multiscale structural disorder associated 
with the complex microscopic texture of the polymer system. \cite{Jerome}

Organic polymer films \cite{Schon} were made to superconduct 
through charge injection in a field-effect-transistor geometry. 
In field-effect transistors, a voltage applied to the gate electrode
controls the charge flow along the surface of the active (conducting)
layer. The active material in the experiment in Ref. \cite{Schon} was 
solution-cast regioregular poly(3-hexylthiophene) (P3HT) deposited on 
an insulating substrate. A superconducting flow was observed below 
$T_c \sim 2.35$ K for the high gate-induced carrier concentration 
of about one charge per 12 thiophene rings. The superconducting 
behavior in P3HT was related to the self-assembly properties of 
the polymer, which forms thin regular films owing to 
self-organization. \cite{Schon,Jerome}

Remark that the gate-induced charge
doping introduces no additional disorder into the material,
enabling a direct observation of the superconducting transition
as a function of temperature and charge carrier concentration.
\cite{Schon,Jerome} Conversely, chemical doping (i.e., injection 
of suitable impurities such as chlorine or alkali metals, leading
to a surplus or deficit of free electrons) inevitably increases 
the disorder. \cite{Schon} This changes the properties of the 
conducting system and can suppress the inherent self-organization 
mechanisms that make organic polymers superconduct. With 
increasing concentration of the impurities, the superconducting
transition temperature $T_c$ shifts to lower values, while the 
transition region becomes generally broader. \cite{Schon} Chemical 
doping has thus defied attempts to turn polymers into superconductors. 
\cite{Schon,Phillips} Until the field-effect experiments in Ref., 
\cite{Schon} the only polymeric material exhibiting superconductivity 
was a crystalline inorganic polymer polysulfur nitride (SN)$_X$ having 
a transition temperature of $\sim 0.26$ K. \cite{Greene}

The key issue on the superconducting organic polymers 
is the comprehension of the electron pairing mechanism. The 
main problem is that electronic properties of the polymers are 
dominated by the intrinsic spatial disorder which supports charge 
localization at low temperatures. \cite{Jerome} (The effect of the 
disorder can be provided, for instance, by a multiscale braiding 
of the polymer molecules.) Superconducting transition in organic 
polymers have been discussed \cite{Jerome} in connection with both 
the conventional picture of Bardeen, Cooper, and Schrieffer (BCS) 
superconductivity relying on the electron-phonon interactions in 
regular crystals, \cite{Cooper} and Little's exciton mechanism, in 
which phonons are replaced by other electrons. \cite{Little} An 
alternative approach could be based on a spin-triplet, even-parity 
superconductivity for disordered, interacting electron systems. 
\cite{Triplet} While strong arguments in favour of the traditional 
explanation in terms of phonon-mediated superconductivity have been 
proposed, \cite{Schon,Jerome} the formation of the Cooper pairs on 
the disordered polymer patterns (revealing no regular monocrystalline 
configuration, contrary to the BCS scenario) remains so far 
generally unclear.

In this paper, we propose a feasible electron pairing mechanism 
for the ``strange" materials like organic polymer P3HT that may 
observe superconducting behavior in spite of the intrinsic structural
disorder. We suggest that electron-electron interactions in such materials 
are mediated by {\it fractons}, quantum oscillations of the multiscale
conducting arrays organized in a fractal lattice. The fracton pairing 
mechanism thereby profits from the complex geometry of the 
effective conducting set. 

It is instructive to emphasize that fracton-mediated interactions 
between the current-carrying particles could be characteristic of the
disordered solids as a whole, rather than specifically of the polymeric
materials. Self-assembling organic polymers are considered in our study 
as a suitable guide to the ``strange" superconducting properties of
complex systems, while the fracton mechanism is not attributed  
only to the polymers. 

We assume, as a general condition, that the material posseses 
complex microscopic organization that supports a well-defined 
branch of fracton excitations. We argue that the fracton modes 
may then effectively confine the current-carrying electrons in 
couples below a certain temperature, thus leading to the occurrence 
of the marginal superconducting fluid state. We demonstrate, furthermore, 
that replacing electrons with ``holes" in the fracton scenario leads 
to the intriguing hints regarding the nature of the high-temperature 
superconductivity in copper oxide materials. The fracton approach 
may thereby indicate that the superconducting properties of organic 
polymers and of copper oxides are both parts of the same problem.

The paper is organized as follows. The basic assumptions of our
study are formulated in Sec. II. A short summary on fracton dynamics
is condensed to Sec. III. The fundamental topological properties of 
percolating fractal lattices giving rise to a well-defined fracton
spectrum, are analyzed in Secs. IV$-$V. The fracton wave function 
is derived in Sec. VI from a fractional kinetic equation in the
harmonic approximation. Based on the results obtained, the 
superconducting transition in self-assembling electron-doped 
organic polymers is discussed in Secs. VII$-$VIII. A case 
for high-temperature superconductivity is analuzed in Sec. IX;
both hole-doped complex systems and copper-oxide compounds are 
considered. Implications of the fracton pairing mechanism on ac 
conduction properties of the strange superconductors above the 
marginal transition temperature $T_c$ are addressed in Sec. X. 
The principal conclusions are given in Sec. XI.

\section{Fractal Geometry of Percolation}

We start with a suitable geometric description of the 
disordered molecular (polymeric) structures. Our basic assumptions 
are, 1) self-organization of organic polymers results in development 
of a multiscale conducting network of molecules, 2) the self-organized 
conducting networks form percolating sets, and 3) these sets are critical, 
i.e., at the threshold of percolation. A percolating network is defined 
as an infinite connected web-like configuration embedded in an Euclidean 
space $E^d$, where $d$ is ambient (integer) dimension; in what follows, 
$d\geq 2$. The term ``percolating" is synonymous with ``infinite 
connected" and guarantees that the molecular network conducts on 
the macroscopic scales. The threshold character means that the 
percolating web, although infinite in size, is minimally developed 
in $E^d$, i.e., observes a sort of a structural extremum. More precisely, 
assume the effective conducting web is composed of multiscale conducting 
links whose concentration in the space $E^d$ is quantified by 
the parameter $0\leq q\leq 1$. \cite{Gefen} For $q\rightarrow 0$, no 
connected structure can exist as the web lacks conducting elements on 
all scales; conversely, for $q\rightarrow 1$, the multiscale conducting 
links densely fill the ambient space. As $q$ continuously grows from 
0 to 1, an infinite connected structure (which occupies only 
a fraction of the embedding space $E^d$) first appears for $q=q_c$,
where $0 < q_c < 1$ is some critical concentration of the links.
(Roughly speaking, $q_c$ is ``of the order of one half".)
The value $q=q_c$ can be identified with the percolation threshold 
in the system, corresponding to a minimal (critical) network which 
conducts on large scales. As $q$ approaches $q_c$, the percolation 
correlation (i.e., pair connectedness) length diverges as 
$\xi \sim |q - q_c| ^{-\nu}$, where $\nu$ is the percolation 
length exponent depending solely on the embedding dimension
$d\geq 2$. For $q > q_c$, the probability to belong to the 
infinite connected network behaves as
$Q _{\infty} (q) \propto (q - q_c) ^{\beta} \propto \xi ^{-\beta / \nu}$,
while dc conductivity goes to zero as
$\Sigma (0) \propto (q - q_c) ^{\mu} \propto \xi ^{-\mu / \nu}$.
Along with the exponent $\nu$, the percolation indices
$\beta$ and $\mu$ completely characterize the structural properties 
of the effective conducting set for $q\rightarrow + q_c$. 
Both $Q _{\infty} (q)$ and $\Sigma (0)$ vanish for $q < q_c$.

The microscopic conducting mechanism operating in the
polymeric materials such as P3HT is supported by the conjugated 
structure of alternating single and double bonds linking the carbon
atoms in their backbone. \cite{Jerome} This bonding structure allows
one electron from each carbon atom to move along the molecule and
become conducting. \cite{Jerome,Conduct} The multiscale braiding of
the polymer molecules gives rise to a complex spatial distribution
of the conducting links, modeled in our study by a percolating 
web-like pattern.

A remarkable property of percolating networks at criticality
(i.e., near the marginal threshold $q=q_c$) is their self-similar 
(fractal) geometry. \cite{Gefen,Feder,Percol,Naka} In a physical system, 
the fractal geometry approximation holds in a wide but finite range of 
spatial scales $\chi$, which vary between the microscopic (``lattice") 
distance $\chi _{\min}\sim a$ and the percolation correlation 
length $\chi _{\max}\sim\xi\gg a$. Given arbitrary length scale 
$a\lesssim\chi\lesssim\xi$, one finds the number of the links 
that belong to the percolating network within the volume 
$\chi ^d \subset E^d$ to be $N _{\chi} \propto \chi ^{d_f}$. 
The quantity $d_f$ is the generalized (Hausdorff) dimension of the 
network which is not larger than $d$, i.e., $d_f \leq d$. 
\cite{Feder,Mandel} Specifically, $d_f = d - \beta / \nu$ 
for $a\lesssim\chi\lesssim\xi$. \cite{Gefen} Note that 
$d_f \geq 1$ for fractal networks admitting connected 
pathes to infinity. 

The value of the Hausdorff dimension $d_f$ is in one-to-one correspondence
with the topological entropy of the fractal network, $S_f$: \cite{PRB}
\begin{equation}
S_f\, =\, 1\, -\, 1/d_f.
\eqnum{1}
\end{equation}
The topological entropy $S_f$ measures the structural disorder in 
complex systems. \cite{Mehaute} For well-ordered conducting polymer 
chains without braiding, $d_f \rightarrow 1$ and $S_f \rightarrow 0$.
Such conditions could be realized at low absolute temperatures
$T\rightarrow 0$. For the multiscale path-connected braided
structures, $d_f > 1$ and $S_f > 0$. Similar to the conventional,
thermodynamic entropy, the topological entropy $S_f$ is a
non-decreasing function of $T$, i.e.,
$\partial S_f / \partial T \geq 0$.
From Eq. (1) one obtains
\begin{equation}
\partial d_f / \partial T\, \geq\, 0.
\eqnum{2}
\end{equation}

At length scales $\chi$ exceeding the correlation 
length $\xi$, the fractal geometry of percolation crosses 
over to a statistically homogeneous distribution of the conducting 
links. This means $N _{\chi} \propto \chi ^{d}$ for $\chi\gtrsim\xi$.
At the short length scales $\chi\lesssim a$, the material may reveal 
regular crystalline structure; in polymers like P3HT, this structure 
can be associated with ordered nanocrystals embedded in a disordered 
matrix. \cite{Jerome} The typical size of the nanocrystals is,
$a\sim 10^{-6}$ cm. \cite{Schon} Note that the ratio $\xi / a$ 
in real fractal materials is at most $\sim 10 ^3$ due to the 
limitations imposed by thermal vibrations and by 
gravitational stability. \cite{Eric} 

The formation of fractal conducting networks in the range 
of scales $a\lesssim\chi\lesssim\xi$ may be explained by 
{\it self-organized criticality} (SOC), a phenomenon which 
often underlies the development of fractal structures in complex 
nonlinear dynamical systems. \cite{Bak} Here, the concept of SOC 
is understood as a synonymous with ``self-organization to a state 
of critical percolation." \cite{PRB,PRE01} The inherent dynamics 
of the ideal SOC state is manifested in the low frequency $f^{-1}$ 
fluctuation spectrum \cite{Bak} customarily referred to as flicker 
noise. The feasible existence of the anomalously large flicker noise 
in resistance fluctuations for disordered solid materials 
was addressed in Ref. \cite{Dyre}.

\section{Fracton Dynamics}

At the quantum level, the dynamics of a percolating fractal network 
can be described in terms of elementary (quasi)acoustic excitations
termed {\it fractons}. \cite{AO} Fractons are considered as
quantum oscillations of the local disorder in the system. In the
harmonic approximation, fractons are well defined quantum states
with infinite inelastic lifetimes. \cite{Orbach} Statistical
properties of fractons were reviewed by Nakayama {\it et al.}
\cite{Naka} in connection with the topological characteristics of
fractal sets. A direct experimental observation of fracton 
modes in real complex materials was reported in Ref. \cite{Eric}

Fracton excitations of the critical percolating
networks form a well-defined (quasi)acoustic branch
representing vibrations of multiscale fractal blobs. \cite{Orbach}
In fractal geometries, such vibrations acquire the role of ordinary
phonons propagating through periodic crystal lattices. (In polymeric 
materials like P3HT, the ``blobs" may be identified with the disordered 
polycrystalline matrices.) Note that phonons having wavelengths shorter 
than the percolation correlation length $\xi$ are strongly scattered by 
the structural inhomogeneitites associated with the fractal geometry of 
the underlying pattern. Hence phonons cannot be eigen vibrational modes 
of a fractal. Conversely, fractons are localized states, in the 
Anderson sense, \cite{And} with a localization length larger than the 
characteristic phonon scattering length. \cite{Eric,Orbach}

The defining feature of the fracton branch is 
the nonlinear relationship \cite{Naka,AO,Orbach}
\begin{equation}
\omega\, \propto\, q ^{\sigma} 
\eqnum{3}
\end{equation}
between angular frequency $\omega$ and wave vector
$q = 2\pi / \lambda$. Here, $\lambda$ is the fracton 
wavelength, which lies between the characteristic fractal 
cutoff scales $\chi _{\min}\sim a$ and $\chi _{\max}\sim\xi$, i.e.,
$a\lesssim\lambda\lesssim\xi$. The power exponent $\sigma$
in the dispersion relation (3) equals
\begin{equation}
\sigma\, =\, (2\, +\, \theta) / 2, 
\eqnum{4}
\end{equation}
where $\theta$ is the index of connectivity of the fractal.
The index of connectivity determines the microscopic composition
of the fractal object through the marginal links between the 
constituent elements. \cite{Procaccia} The value of $\theta$
does not depend on the way the fractal is folded in the embedding
Euclidean space $E^d$ (as opposed to the Hausdorff dimension $d_f$ 
which relies on the embedding properties of the fractal). Contrary
to the Hausdorff dimension $d_f$, the index of connectivity $\theta$ 
is a topological invariant of the fractal structure, i.e., remains 
unchanged under the homeomorphic deformations that preserve the 
property of self-similarity. \cite{PRE00} The index of connectivity
$\theta$ can be expressed in terms of the percolation exponents
$\nu$, $\beta$, and $\mu$ via $\theta = (\mu - \beta) / \nu$. 
\cite{Gefen} In Euclidean geometries, $\theta\equiv 0$.

The fracton phase velocity, $c_{\sigma}$, scales with wave vector 
$q$ as $c _{\sigma} (q) \sim \omega / q \propto q^{\theta / 2}$. For 
$q\xi / 2\pi \lesssim 1$, the fracton branch crosses over to the 
long-wave phonons, whose phase velocity, $c$, is a constant function 
of $q$, i.e., $c \sim \omega / q =$ const $(q)$. \cite{Naka,Orbach} 
Note that phonons are characterized by the linear dispersion relation 
$\omega\propto q$, which follows from Eq. (3) in the Euclidean
geometry limit $\theta\rightarrow 0$. In the fractal domain, we
have, consequently,
\begin{equation}
c _{\sigma} (q)\, \sim\, c \times (q\xi / 2\pi) ^{\theta / 2},
\eqnum{5}
\end{equation}
where the fracton wave vector $q$ varies from 
$q_{\min} \sim 2\pi / \xi$ to $q_{\max} \sim 2\pi / a$. 
Considering Eq. (5), one concludes that the fracton phase 
velocity cannot exceed the maximum of 
\begin{equation}
c _{\sigma\, \max}\, \sim\, c \times (\xi / a) ^{\theta / 2},
\eqnum{6}
\end{equation}
corresponding to the shortest-wave fractons with $q \sim q_{\max}$. 
Accordingly, the fracton frequecies range through
\begin{equation}
\omega_{\min}\,\lesssim\,\omega\,\lesssim\, 
\omega_{\min} \times (\xi / a) ^{\sigma},
\eqnum{7}
\end{equation}
where $\omega _{\min} \sim 2\pi c / \xi$ 
is the long-wave phonon crossover scale.

The fracton density of states (DOS) behaves
with frequency $\omega$ as \cite{Naka,AO,Orbach}
\begin{equation}
{\cal {D}} _{\textit{\emph{fr}}} (\omega)\, \propto\, \omega ^{d_s - 1},
\eqnum{8}
\end{equation}
where $d_s = 2 d_f / (2 + \theta) = d_f / \sigma$ is the so-called 
spectral fractal dimension \cite{AO} (to be distinguished from the 
Hausdorff dimension of the fractal, $d_f$). The value of $d_s$
determines the effective (fractional) number of degrees 
of freedom on a fractal geometry. \cite{Georges} 
In the limit $d_s\rightarrow d$, expression (8) 
crosses over to the phonon DOS \cite{Feynman}
\begin{equation}
{\cal {D}} _{\textit{\emph{ph}}} (\omega)\, \propto\, \omega ^{d - 1}.
\eqnum{9}
\end{equation}

\section{The Percolation Constant}

Near the threshold of percolation, $q=q_c$, 
the value of the spectral dimension $d_s$ can 
be obtained from topological arguments. \cite{PRE97}
In fact, let $\partial {\cal {N}} _c$ and ${\cal {N}} _c$ denote the
critical and fat percolating networks, respectively. The fat percolating
network is defined as the set of points which can be path connected
\cite{Fomenko} to $\partial {\cal {N}} _c$ for all $q < q_c$. Obviously,
${\cal {N}} _c$ is an open set, \cite{Topol} and $\partial {\cal {N}} _c$
is the boundary \cite{Topol} of ${\cal {N}} _c$. (This motivates the
commonly used sign ``$\partial$" in the notation 
$\partial {\cal {N}} _c$.) We also define the closed set \cite{Topol} 
$\bar {{\cal {N}}} _c = {\cal {N}} _c \bigcup \partial {\cal {N}} _c$.

It can be proven \cite{PRE97} that
$\partial {\cal {N}} _c$ and ${\cal {N}} _c$ are
topologically equivalent with the fractional $(d_s - 1)$-dimensional 
sphere $S^{d_s - 1}$ and the fractional $d_s$-dimensional open disk 
$D^{d_s}$, respectively, where $S^{d_s - 1} = \partial D^{d_s}$ is 
the boundary of $D^{d_s}$. The introduction of the {\it fractional 
manifolds} \cite{PRE97} $S^{d_s - 1}$ and $D^{d_s}$ extends the notion 
of the smooth integer-dimensional manifolds \cite{Fomenko,Modern} 
$S^{d - 1}$ and $D^{d}$ to arbitrary real dimensionalities $d_s < d$.
Note that the fractional sphere $S^{d_s - 1}$ is based on the solid angle
\begin{equation}
\eqnum{10}
\Omega _{d_s}\, =\, d_s \frac
{\pi ^{d_s /2}}
{\Gamma (d_s /2 + 1)},
\end{equation}
where $\Gamma$ is the Euler gamma-function.
From Eq. (10) one recovers the familiar results
$\Omega _2 = 2\pi$ for the standard circle $S^1$, and
$\Omega _3 = 4\pi$ for the standard two-dimensional sphere $S^2$.
The topological equivalence relation can be given by a diffeomorhism
$\{ \phi:~\partial {\cal {N}} _c \rightarrow S^{d_s - 1},~
{\cal {N}} _c \rightarrow D^{d_s} \}$. (Here, the diffeomorphic 
relation $\phi$ implies fractional extension \cite{PRE97} of the 
Jacobian \cite{Modern}.)

As is shown in Ref. \cite{PRE97}, the closed fractional manifold
${\bar D}^{d_s} = D^{d_s} \bigcup S^{d_s - 1}$ admits everywhere a
dense covering by a fractal curve $\gamma ^*$ which has {\it no} points
of self-crossings at the critical threshold $q = q_c$. The absence of
points of self-crossings minimizes the topology of the network for
$q=q_c$ and thereby supports the conditions of the critical percolation.
(For $q > q_c$, the corresponding manifolds contain ``too many points"
to be densely covered by a curve without self-crossings. The minimal closed
percolating structure $\bar {{\cal {N}}} _c \subset E^d$ therefore contains 
``the same number of points" as a fractal curve without self-crossings, 
folded in $E^d$. This serves as the defining feature of the critical 
percolation. \cite{PRE97}) Because $\gamma ^*$ is dense everywhere in 
${\bar D}^{d_s}$, it must be based on exactly the same solid angle 
$\Omega (\gamma ^*) = \Omega _{d_s}$ as the
fractional sphere $S^{d_s - 1} = \partial D^{d_s}$ [see Eq. (10)].
On the other hand, a curve $\gamma ^*$ cannot be principally based
on a solid angle $\Omega (\gamma ^*)$ smaller than $\Omega _2 /2 = \pi$
in $d\geq 2$ embedding dimensions, i.e., $\Omega (\gamma ^*) \geq \pi$
for $d\geq 2$. The criticality condition $q = q_c$ minimizes
the solid angle $\Omega _{d_s}$, i.e.,
$\Omega (\gamma ^*) = \Omega _{d_s} = \pi$ at the
percolation threshold. \cite{PRE97} The solution to this
transcendental algebraic equation is the fundamental topological
parameter ${\cal {C}}$ (termed ``the percolation constant" in Ref. 
\cite{PRE00}) satisfying, by definition, the identity \cite{PRE97}
\begin{equation}
\eqnum{11}
{\cal {C}} \frac
{\pi ^{{\cal {C}} /2}}
{\Gamma ({\cal {C}} /2 + 1)}\, =\, \pi.
\end{equation}
Numerically, ${\cal {C}}\approx 1.327$. \cite{PRE97} Hence, the spectral
fractal dimension
\begin{equation}
\eqnum{12}
d_s\, \equiv\, 2 d_f / (2\, +\, \theta)\, =\, {\cal {C}}\, \approx\, 1.327
\end{equation}
at $q = q_c$. Based on the Whitney embedding theorem, \cite{Whitney}
one may demonstrate \cite{PRE97} that Eq. (12) holds for embedding
dimensions $d$ between 2 and 5, i.e., $2\leq d\leq 5$. Note that the
value of $d_s$ at $q=q_c$ is given by the mean-field theory \cite{Mean}
for all $d\geq 6$: $d_s = 4/3$. Below the critical embedding
dimension of $d=6$, the mean-field percolation is invalidated;
\cite{Naka,Mean} this appears in the deviation of the percolation
constant ${\cal {C}}$ from $4/3$ for $2\leq d\leq 5$. The fact that
${\cal {C}}$ is actually {\it smaller} than $4/3$ is in accord with
the prediction of the renormalization-group $\epsilon$-expansion
to the marginal orders in $\epsilon$. \cite{Naka,Mean} The issue 
of the percolation constant ${\cal {C}}$ \cite{PRE00,PRE97} is 
intimately connected to the widely-known {\it Alexander-Orbach 
conjecture}, \cite{AO} which up to recently has been amongst the 
biggest open challenges in the fundamental percolation theory. Note, 
also, that Eq. (12) is restricted to contractible \cite{Fomenko} 
fractal sets; the noncontractibility effects tend to diminish the 
actual value of the spectral fractal dimension $d_s$ at $q=q_c$ 
when compared to the percolation constant ${\cal {C}}\approx 1.327$. 
\cite{PRE00,PRE97}

In view of Eq. (12), the fracton DOS near criticality becomes
\begin{equation}
{\cal {D}} _{\textit{\emph{fr}}} (\omega)\, 
\propto\, \omega\,^{{\cal {C}} - 1}\,
\sim\, \omega ^{1/3},
\eqnum{13}
\end{equation}
where the mean-field estimate ${\cal {C}}\sim 4/3$ 
has been used for simplicity. From Eqs. (8), (9) and (13) 
one concludes that the fracton states ($d_s = {\cal {C}} \sim 4/3$) 
are much denser than the phonon states ($d_s = d\geq 2$) for the
frequencies $\omega$ smaller than $\sim \omega_{\min} \times 
(\xi / a) ^{\sigma}$. Fractons could thereby dominate in the 
excitation spectrum of the system below the 
characteristic temperature of 
\begin{equation}
T_c\,\sim\, \hbar \omega_{\min} \times (\xi / a) ^{\sigma}\, \sim\,
2\pi \hbar (c / a) \times (\xi / a) ^{\theta / 2},
\eqnum{14}
\end{equation}
in accordance with condition (7). 

\section{Topological Constraints on the Hausdorff Dimension}

In this section, topological constraints on the Hausdorff
dimension $d_f$ are found. The results obtained are further used 
to define the admissible values of the index of connectivity $\theta$ 
and of the exponent $\sigma = (2 + \theta) / 2$ in Eq. (14).

First, let us observe that the closed fat percolating network
${\bar {\cal {N}}} _c = \phi ^{-1} {\bar D}^{d_s}$ can be densely 
covered everywhere by a fractal curve $\Gamma ^* = \phi ^{-1} \gamma ^*$
which has no points of self-crossings. (Remind the diffeomorphism 
$\{ \phi:~\partial {\cal {N}} _c \rightarrow S^{d_s - 1},~
{\cal {N}} _c \rightarrow D^{d_s} \}$ between the closed fat percolating
network ${\bar {\cal {N}}} _c = {\cal {N}} _c \bigcup \partial {\cal {N}} _c$ 
and the fractional manifold ${\bar D}^{d_s} = D^{d_s} \bigcup S^{d_s - 1}$.) 
We now prove that the Hausdorff fractal dimension, $\dim \Gamma ^*$, of 
the curve $\Gamma ^*$ is equal to $d_f \equiv \dim \partial {\cal {N}} _c$, 
\cite{PRB} i.e.,
\begin{equation}
d_f\, \equiv\, \dim \partial {\cal {N}} _c\, =\, \dim \Gamma ^*.
\eqnum{15}
\end{equation}
In fact, because $\Gamma ^*$ is dense everywhere in ${\bar {\cal {N}}} _c$,
one has $\dim \Gamma ^* = \dim {\bar {\cal {N}}} _c$. We need to demonstrate
that the critical percolating network $\partial {\cal {N}} _c$ is dense 
everywhere in the closed fat percolating network ${\bar {\cal {N}}} _c$: 
This would imply $\dim \partial {\cal {N}} _c = \dim {\bar {\cal {N}}} _c$,
thus leading to Eq. (15).

{\it Lemma:}~$\partial {\cal {N}} _c$
{\it is dense everywhere in} ${\bar {\cal {N}}} _c$.
This means, precisely, that for arbitrary $\varepsilon > 0$ and
arbitrary point $A_c \in {\cal {N}} _c$ one finds another point
$A^*_c \in \partial {\cal {N}} _c$, such that the distance
$||A_c - A^*_c|| \leq \varepsilon$. \cite{Topol} [Roughly speaking, the 
(open) fat percolating network ${\cal {N}} _c$ has an ``extremely developed"
boundary $\partial {\cal {N}} _c$ at $q = q_c$.] Note that $||\dots ||$
is Euclidean distance in the embedding $d$-dimensional space. Assume the
contrary: $\partial {\cal {N}} _c$ is {\it not} dense everywhere in
${\bar {\cal {N}}} _c$. Hence there exist $\varepsilon > 0$
and $A_c \in {\cal {N}} _c$, such that $A_c$ can be surrounded
by the closed $(d\geq 2)$-dimensional Euclidean $\varepsilon$-disk
${\bar D}_{\varepsilon}^{d}$ which lies in ${\cal {N}} _c$
but contains no points of the boundary $\partial {\cal {N}} _c$, i.e.,
${\bar D}_{\varepsilon}^{d} \subset {\cal {N}} _c$,
${\bar D}_{\varepsilon}^{d} \bigcap \partial {\cal {N}} _c = \emptyset$.
Because $\Gamma ^*$ densely covers ${\bar {\cal {N}}} _c$
everywhere, the $\varepsilon$-element
$\Gamma _{\varepsilon} ^* = \Gamma ^* \bigcap {\bar D}_{\varepsilon}^{d}$
densely covers ${\bar D}_{\varepsilon}^{d}$ everywhere for $d\geq 2$.
Remark that $\Gamma _{\varepsilon} ^*$ has no points of self-crossings.
Consequently, we constructed a dense covering everywhere of a
($d\geq 2$)-dimensional Euclidean domain ${\bar D}_{\varepsilon}^{d}$
by a fractal curve $\Gamma _{\varepsilon} ^*$ which has no
points of self-crossings. This, however, is principally impossible
in $d\geq 2$ dimensions. \cite{Nagltop} For instance, a dense covering 
everywhere of a two-dimensional ($d=2$) Euclidean domain
${\bar D}_{\varepsilon}^{2}$ is provided
by the Peano curve which has infinite number of points of
self-crossings (e.g., of multiplicity 4) where different segments of the
curve are in contact with each other. \cite{Nagltop} The observed
contradiction proves the lemma.

The next step is to consider general embedding conditions for the
curves that have no points of self-crossings. Indeed, a plane ($d=2$)
curve can be embedded without self-crossings into at most
${\cal {S}} _2 = \ln 8 / \ln 3 \approx 1.89 < 2$ dimensions, 
\cite{Nagltop} where ${\cal {S}} _2 \equiv \ln 8 / \ln 3$ is 
the Hausdorff fractal dimension of the square Sierpinski carpet. 
\cite{Feder,Schroeder} (The inequality ${\cal {S}} _2 < 2$ excludes 
existence of the Peano curves without self-crossings.) Analogously, a 
spatial ($d=3$) curve can be embedded without self-crossings into no 
more than ${\cal {S}} _3 = \ln 26 / \ln 3 \approx 2.96 < 3$ dimensions, 
\cite{Nagltop,Schroeder} where ${\cal {S}} _3 \equiv \ln 26 / \ln 3$ is 
the Hausdorff fractal dimension of the three-dimensional Cantor cheese.
The existence of an embedding into the Sierpinski carpet (Cantor cheese)
for the given plane (spatial) curve is a topologically invariant property
\cite{Nagltop} that does not depend on the way the curve is folded in space.
An application of this property to the fractal curve $\Gamma ^*$ leads to
the following conditions: $1\leq \dim \Gamma ^* \leq {\cal {S}} _2$ for 
$d=2$; $1\leq \dim \Gamma ^* \leq {\cal {S}} _3$ for $d=3$. In arbitrary
dimensions $d\geq 2$, one gets, accordingly
\begin{equation}
1\, \leq\, \dim \Gamma ^*\, \leq\, {\cal {S}} _d\, \equiv\,
\ln (3^d\, -\, 1) / \ln 3\, <\, d,
\eqnum{16}
\end{equation}
where ${\cal {S}} _d \equiv \ln (3^d - 1) / \ln 3$ is the 
Hausdorff dimension of the Cantor cheese in $E^d$. \cite{Schroeder} 
Considering Eq. (15), one finds
\begin{equation}
1\, \leq\, d_f\, \leq\, {\cal {S}} _d\, \equiv\,
\ln (3^d\, -\, 1) / \ln 3,
\eqnum{17}
\end{equation}
where $2\leq d\leq 5$ due to the limitations imposed
by the Whitney embedding theorem. \cite{PRE97} Inequality 
(17) establishes the admissible values of the Hausdorff 
dimension $d_f$ for percolating networks at criticality
($2\leq d\leq 5$). From Eqs. (12) and (17) we finally obtain
\begin{equation}
- 2({\cal {C}} - 1) / {\cal {C}}\, \leq\, \theta\, 
\leq\, 2({\cal {S}} _d - {\cal {C}}) / {\cal {C}},
\eqnum{18}
\end{equation}
\begin{equation}
1 / {\cal {C}}\, \leq\, \sigma\, \leq\, {\cal {S}} _d / {\cal {C}}.
\eqnum{19}
\end{equation}
Inequalities (18)$-$(19) hold for embedding dimensions $2\leq d\leq 5$.
From Eqs. (2) and (12) one also gets
\begin{equation}
\partial \theta / \partial T\, = 2\, \partial \sigma / \partial T\, \geq\, 0,
\eqnum{20}
\end{equation}
provided the system is at the threshold of percolation.

\section{Fractional Wave Equation for Fracton Excitations}

We now turn to a calculation of the fracton wave function, $\Psi (t, x)$,
in a quantum state with angular frequency $\omega$ and wave vector $q$. In 
the harmonic approximation, we have $\Psi (t, x) = \psi (x) e ^{i\omega t}$.
The real-space coordinate $x$ is chosen in the direction of $q$. 
The standard normalization condition
\begin{equation}
\int _{-\infty}^{+\infty}\, |\Psi (t, x)| ^2 dx\, =\, 
\int _{-\infty}^{+\infty}\, |\psi (x)| ^2 dx\, =\, 1
\eqnum{21}
\end{equation}
is implied. The Planck's constant $\hbar = 1$ in this section.

We start from the fracton dispersion 
relation $\omega = \Lambda q ^{\sigma}$ with 
$\Lambda \sim c \times (\xi / 2\pi) ^{\theta / 2}$, 
in accordance with Eqs. (3) and (5). Replacing $\omega$ 
and $q$ by the operators $i\partial / \partial t$ and 
$-i\nabla\equiv -i\partial / \partial x$, respectively, 
and taking into account the harmonic form 
$\Psi (t, x) = \psi (x) e ^{i\omega t}$, 
one arrives at the fractional kinetic equation 
\begin{equation}
i\partial\Psi (t, x) / \partial t\, =\, i^{-\sigma}
\Lambda \nabla ^{\sigma} \Psi (t, x),
\eqnum{22}
\end{equation}
where $\nabla ^{\sigma}$ denotes the $\sigma$-th power of 
$\nabla$ (to be considered as the fractional generalization of 
the Laplacian, $\Delta\equiv \nabla ^2$). An explicit analytical 
representation of the fractional derivative $\nabla ^{\sigma}$ 
is given by the Riesz operator defined in Ref. \cite{Klafter} 
The introduction of the fractional Eq. (22) for the (quasi)acoustic
excitations in fractal networks follows the paradigm of the {\it strange}
(or {\it fractional}) dynamics \cite{Nature} which has come of age as a 
complementary tool in the description of anomalous kinetic processes 
in complex systems. \cite{Klafter,PhysicaD,Today,PhysicaA,Plasma,PRE0101}  

The real-space counterpart, $\psi (x)$, of the fracton wave function
$\Psi (t, x) = \psi (x) e ^{i\omega t}$ obeys the fractional equation 
\begin{equation}
\omega \psi (x)\, =\, -\,i^{-\sigma}
\Lambda \nabla ^{\sigma} \psi (x).
\eqnum{23}
\end{equation}
General analytical solution to Eq. (23) can be expressed in terms
of Mittag-Leffler functions (these being a special case of  
Fox functions \cite{Klafter}). Elementary-function solutions may
be obtained in the following limiting cases:

1)~{\it Core region:}~$\omega x ^{\sigma} \lesssim \Lambda$:
\begin{equation}
\psi _n (x)\, \sim\, \exp 
\left(
-\,i ^{\sigma} \frac
{\omega x ^{\sigma}}
{\Lambda \Gamma (1 + \sigma)}
\right),
\eqnum{24}
\end{equation}
\begin{equation}
\Psi _n (t, x)\, \sim\, e ^{i\omega t}\, \exp
\left(
-\,i ^{\sigma} \frac
{\omega x ^{\sigma}}
{\Lambda \Gamma (1 + \sigma)}
\right),
\eqnum{25}
\end{equation}
\begin{equation}
|\Psi _n (t, x)| ^2\, =\, |\psi _n (x)| ^2\, \sim\, \exp 
\left(
- \frac
{2\omega\alpha _n x ^{\sigma}}
{\Lambda \Gamma (1 + \sigma)}
\right),
\eqnum{26}
\end{equation}
where $i ^{\sigma} \equiv \cos [\pi\sigma (4n + 1) / 2] 
+ i \sin [\pi\sigma (4n + 1) / 2]$, $n = 0, \pm 1, \pm 2, \dots$ is 
an integer number, and $\alpha _n = \cos [\pi\sigma (4n + 1) / 2]$
is the real part of $i ^{\sigma}$. Admissible (decaying) fracton modes 
$\Psi _n (t, x)$ are selected by the condition 
\begin{equation}
\alpha _n\, =\, \cos [\pi\sigma (4n + 1) / 2]\, \geq\, 0.
\eqnum{27}
\end{equation}
The shape of the fracton wave function in the core region
$\omega x ^{\sigma} \lesssim \Lambda$ is thereby a stretched
exponential distribution $\psi _n (x) \sim \exp 
[- (x / \lambda _n) ^{\sigma}]$, with
$\lambda _{n} ^{\sigma} = \Lambda\Gamma 
(1 + \sigma) / \omega\alpha _n$. The spatial
scale $\lambda _n$ (i.e., the size of the core region) is 
often referred to as the fracton localization length. \cite{Orbach} 
Note that the values of $\sigma$ lie in the interval (19) and may 
depend on the embedding dimension $2\leq d\leq 5$ in general.
Numerically, $0.75\lesssim \sigma\lesssim 1.42$ for $d = 2$, 
and $0.75\lesssim \sigma\lesssim 2.23$ for $d = 3$. In Euclidean 
geometries, $\sigma = 1$, yielding $\alpha _n \equiv 0$ for all 
$n$. In this limit, the fracton wave function $\Psi _n (t, x)$ 
crosses over to a plane wave, $\sim\exp (i\omega t - iqx)$.

The stretched exponential behavior of the fracton wave function
in the core region was first proposed by Entin-Wohlman {\it et al.}
\cite{Entin} Much theoretical and numerical effort has later been 
invested to determine the value of $\sigma$ (for a review, see 
Nakayama {\it et al.}, \cite{Naka} and references therein. 
In the notations of article, \cite{Naka} $\sigma\equiv d_{\phi}$.) 
The results addressed in Ref. \cite{Naka} comply well with 
the constraint (19) in both cases $d=2$ and $d=3$. 

2)~{\it Tail region:}~$\omega x ^{\sigma} \gg \Lambda$:
The asymptotic ($\omega x ^{\sigma} \gg \Lambda$) analysis 
of the fractional Eq. (23) is similar to a derivation of the
L\'evy stable distributions \cite{Klafter} from the Riesz
operator. Making use of the normalization condition (21),
from Eq. (23) one finds [cf. Eqs. (24)$-$(26)] 
\begin{equation}
\psi _n (x)\, \sim\,
\left(
\frac
{\Lambda}
{\omega\Gamma (-\sigma)}
\right) ^{1/2}\,\times\,
x ^{- (\sigma + 1) /2},
\eqnum{28}
\end{equation}
\begin{equation}
\Psi _n (t, x)\, \sim\, 
\left(
\frac
{\Lambda}
{\omega\Gamma (-\sigma)}
\right) ^{1/2}\, e ^{i\omega t}\, \times\,
x ^{- (\sigma + 1) /2},
\eqnum{29}
\end{equation}
\begin{equation}
|\Psi _n (t, x)| ^2\, =\, |\psi _n (x)| ^2\, \sim\,
\frac
{\Lambda}
{\omega |\Gamma (-\sigma)|}\,\times\,
x ^{- (\sigma + 1)}.
\eqnum{30}
\end{equation}
Equation (30) represents the power-law tail 
of the fracton wave function $\Psi _n (t, x)$, 
with the exponent $\sigma + 1$ varying from approximately 
1.75 to $\approx$ 2.42 in two dimensions ($d=2$), and up to 
$\approx$ 3.23 in three dimensions ($d=3$). These relatively
small values of the power exponent $\sigma + 1 \lesssim 3.5$ allow 
for the considerable probabilities of finding a fracton beyond the 
core region when compared to a stretched exponential interpolation. 
We emphasize that the power-law tail (30) is the fractional dynamics 
property contained in the Riesz operator $\nabla ^{\sigma}$. 

\section{The Interaction Hamiltonian}

Similar to the ordinary phonons in regular crystals,
fractons may mediate electron-electron interactions in
fractal media. The implication is that an electron placed 
on a fractal network generates a structural deformation which
affects another electron. The process is described by the 
interaction Hamiltonian 
\begin{equation}
{\cal {H}}\, =\, \sum\limits _{{\bf k}, {\bf p}}\,
{\cal {W}} _{{\bf k} {\bf p}} \phi _{{\bf {p}-{\bf k}}}
a^+_{{\bf p}} a_{{\bf k}}\, +\,
\sum\limits _{{\bf k}, {\bf p}}\,
{\cal {W}}\, ^* _{{\bf k} {\bf p}} \phi ^+ _{{\bf {p}-{\bf k}}}
a^+_{{\bf k}} a_{{\bf p}},
\eqnum{31}
\end{equation}
where $a^+ _{\bf k}$ is the creation operator for an electron
with wave vector ${\bf k}$, $\phi ^+ _{{\bf {p}-{\bf k}}}$ is the
creation operator for a fracton with wave vector ${\bf {p}-{\bf
k}}$, and ${\cal {W}} _{{\bf k} {\bf p}}$ are the elements of 
the interaction matrix, maximized for ${\bf p} = {\bf - k}$. 
The fracton operators $\phi ^{+} _{{\bf {p}-{\bf k}}}$
and $\phi _{{\bf {p}-{\bf k}}}$ are introduced for the 
excitations that obey the selection rule (27). The Hamiltonian in Eq. 
(31) recovers the conventional electron-phonon interaction Hamiltonian 
\cite{Feynman} in the Euclidean geometry limit $d_f\rightarrow d$ 
and $d_s / d_f \rightarrow 1$. The existence of the electron
coupled states for the Hamiltonian (31) can be demonstrated following
the standard formalism. \cite{Cooper,Feynman} Such states would
physically correspond to the ``Cooper pairs" on fractal networks. 
Like in the conventional, phonon-mediated picture of electron-electron 
interactions, \cite{Cooper,Feynman} the Cooper pairs in fractal
geometry occur because electrons with nearly opposite momenta
$({\bf k},{\bf -k})$ exchange a fracton excitation. 

Owing to the effect of the fractional dynamics, the 
electron-fracton interactions are not restricted to the 
core region $\omega x ^{\sigma} \lesssim \Lambda$, but are 
extended, by means of the Riesz operator $\nabla ^{\sigma}$, 
to the ``strange" domain $\omega x ^{\sigma} \gg \Lambda$
corresponding to the power-law tail of the fracton wave function. 
The tail counterpart given by Eqs. (28)$-(30)$ supports broad 
electron-electron couplings in the medium, enabling the formation 
of the Cooper pairs throughout the self-similarity interval 
$a\lesssim\chi\lesssim\xi$. In view of inequality (20), the 
couplings are more extensive at lower absolute temperatures $T$. 
In view of the high density of states in Eq. (13) (large compared 
to the phonon DOS), fracton excitations beneficiate as a 
favourable ingredient of the superconducting transition 
in strange materials.

\section{Tunneling Effects}

In fractal networks, the onset of superconductivity may be 
limited to the Anderson localization of the Cooper pairs. In 
fact, as it follows from inequalities (2) and (20), the fractal 
geometry of the network creates a potential barrier for a particle, 
with the height proportional to $T$. A superconducting behavior on a 
network structure could be the case if the quantum tunneling length 
for the charge carriers exceeds the characteristic Anderson localization
length on the fractal. The Anderson localization length, in its turn,
is typically of the order of the microscopic distance $a$, describing
the characteristic scale of the inhomogeneities present. The quantum 
tunneling length for a particle of mass $m^*$ is given by the well 
known estimate $l^* \sim \hbar / \sqrt{2m^* T}$, where $T$ stands 
for the height of the barrier. A transition to a superconducting 
state may thereby occur if $l^*\gtrsim a$. Taking into account 
that $m^* = 2m_e$ for the Cooper pairs, one finds the 
appropriate temperatures to be
\begin{equation}
T\, \lesssim\, \hbar ^2 / 4m_e a^2.
\eqnum{32}
\end{equation}
(We imply that the material is well supplied with the free
electrons that may effectively participate in the pairing
processes. In the experiments in Ref., \cite{Schon} the necessary 
charge carrier concentration was achieved through the gate-induced 
doping in the field-effect device.) Setting $a\sim 10^{-6}$ cm, 
\cite{Schon} we have $T \lesssim$ (2$-$3) K. This complies with 
the onset temperature $T_c \sim 2.35$ K reported in Ref. 
\cite{Schon} 

The tunneling of the Cooper pairs through braided 
molecular structures is the marginal dynamical process that 
limits the superconducting behavior in the polymeric systems. 
The corresponding temperatures $T \lesssim$ (2$-$3) K are rather
modest owing to the relatively large threshold tunneling length 
$l^*\sim a \sim 10^{-6}$ cm on the right of inequality (32). 
[Decreasing this length might promise enhanced temperatures,
of course.] Anticipating the results of the next Section, we 
note that the fracton-mediated superconductivity may come 
into play at the considerably higher temperatures (typically, 
as high as $T_c \sim 150$ K), provided the tunneling limitation
(32) is circumvented. 

The tunneling processes on a fractal set could be 
associated with the formation of a disordered array of
Josephson junctions enveloping the domains of effective
electron-fracton interactions. In this context, it is worth
noting that superconductivity, due to Josephson coupling of the
superconducting domains, was in fact observed in polycrystalline
systems. \cite{Joseph} The role of Josephson arrays for the
superconducting behavior in P3HT was addressed in Ref. \cite{Schon}
in connection with the patchy (granular) topology of
the transition regions.

The fracton pairing mechanism might explain why superconductivity has
not been observed in inorganic semiconductor devices, \cite{Phillips}
such as Si and GaAs: The active materials used in such devices lack 
the structural complexity which might support a well-defined branch 
of fracton excitations in the self-similar domain. Conversely, 
self-assembling organic polymers like P3HT reveal a complex 
organization owing to SOC, that offers a fertile playground 
for the fracton modes as soon as the criticality condition 
(12) is satisfied. 

\section{A Case for High-Temperature Superconductivity}

An important ramification of the present study 
is a model when electrons are replaced by charged 
``holes" participating in the fracton exchange processes 
on a complex substrate. The simplest interaction Hamiltonian reads
\begin{equation}
{\cal {H}}\, =\, \sum\limits _{{\bf k}, {\bf p}}\,
{\cal {W}} _{{\bf k} {\bf p}} \phi _{{\bf {p}-{\bf k}}}
b^+_{{\bf p}} b_{{\bf k}}\, +\,
\sum\limits _{{\bf k}, {\bf p}}\,
{\cal {W}}\, ^* _{{\bf k} {\bf p}} \phi ^+ _{{\bf {p}-{\bf k}}}
b^+_{{\bf k}} b_{{\bf p}}.
\eqnum{33}
\end{equation}
Here, $b ^{+} _{\bf k}$ and 
$b _{\bf k}$ denote the creation-annihilation 
operators for a hole with wave vector ${\bf k}$. The expression
(33) is analogous to the Hamiltonian in Eq. (31) for the fracton-mediated 
electron coupling. The interaction processes (33) could dominate in 
the nominally insulating materials at high hole-doping levels. 
Beside hole-doped systems, the hole coupling effects might be 
essential in cuprate superconductors owing to the specific 
crystalline structure of the copper-oxide compounds. In fact, 
the cuprate superconductors consist of parallel planes of copper 
and oxygen atoms arranged in a square grid. The copper-oxygen 
planes are separated by the layered atoms of other elements, which 
may absorb electrons from the copper sites, leaving positively charged 
holes behind. \cite{Focus} The integral system operates like a multi-layer
field-effect transistor with the conducting copper-oxygen planes confined
between the charge-absorbing insulating substrates. The complex microscopic
texture of the compound supports fractional harmonic modes in Eq. (22) 
that may turn the holes into coupled states. 

The key point about the holes is that they have negative mass 
$m^* = - |m^*| < 0$, leading to the imaginary fractal tunneling 
length $l^* \sim i \hbar / \sqrt{2|m^*| T}$. This means periodic
charge oscillations $\sim \exp (-i x \sqrt{2|m^*| T} / \hbar)$
in the copper-oxygen lattice, i.e., a ``stripe". \cite{Focus} The 
stripes form a runway \cite{Focus} along which the coupled holes 
may flow without resistance. The stripe order can be associated with 
self-organization of the system to a thermodynamically 
profitable one-dimensional charge distribution.

It is fair to remark that the issue of stripes is at present 
widely debated in the literature. Theoretical arguments in favour 
of the stripes have been proposed in Ref. \cite{Stripes} Concerns 
come from the poor observational support for the charge stripes as 
an integral part of the superconducting behavior in the cuprates. 
Intriguing experimental indications of the charge-stripe order 
in high-temperature superconductors have been obtained only 
recently. \cite{Evidence1,Evidence2} (See, also, Ref. 
\cite{Poor} where an evidence for the stripes in a 
chemically doped cuprate compound was reported.)

Inclusion of stripes 
%%%% $\sim \exp (-i x \sqrt{2|m^*| T} / \hbar)$ 
makes it possible to bypass the marginal limitation (32) on the
superconducting temperature range. The typical superconducting 
transition temperature $T_c$ can then be directly evaluated from 
expression (14), provided the charge is carried by the fracton 
coupled holes along the stripe runways. Considering condition 
(20), we maximize the index of connectivity $\theta$ in the 
scaling $T_c \propto (\xi / a) ^{\theta / 2}$ to be
$\theta = \theta _{\max} = 2({\cal {S}} _d - {\cal {C}}) / {\cal {C}}$, 
where ${\cal {S}} _d$ is the Hausdorff dimension of the Cantor
cheese \cite{Feder,Schroeder} in $E^d$, and ${\cal {C}}$ is the 
percolation constant. \cite{PRE00,PRE97} (High-temperature regime 
is implied, enabling one to set $\theta = \theta _{\max}$ from
$\partial \theta / \partial T\geq 0$.) In view of the plane ($d=2$) 
geometry of the conducting layers in cuprate superconductors, 
\cite{Focus} we find for the scaling exponent, $\theta / 2 = 
\theta _{\max} / 2 = ({\cal {S}} _2 - {\cal {C}}) / {\cal {C}} 
\approx 0.43$, where ${\cal {S}} _2 \approx 1.89$ is the Hausdorff 
dimension of the square Sierpinski carpet. \cite{Feder,Schroeder} 
Assuming further the maximal \cite{Eric} ratio $\xi / a \sim 10 ^3$, 
we estimate the factor $(\xi / a) ^{\theta / 2}$ in Eq. (14) to be 
around 20. After all, making use of the characteristic values 
$c\sim 2\times 10 ^5$ cm/s for the long-wave phonon phase velocity, 
and $a\sim 10 ^{-6}$ cm for the minimal fracton wavelength, we have, 
by order of magnitude, $T_c \sim 40 \pi \hbar c / a \sim 150 K$. This 
result is in good agreement with the typical transition temperatures 
observed in cuprate superconductors. Enhanced transition
temperatures might be speculated for the compounds having
$2\pi c / a \gtrsim 10^{12}$ Hz.

We emphasize that the high 
transition temperature $T_c \sim 150$ K 
profits from the imaginary hole tunneling 
length, $l^* \sim i \hbar / \sqrt{2|m^*| T}$, 
giving rise to the self-organized charge-stripe 
order across the conducting set. In cuprate superconductors, 
this macroscopic order helps coupled holes run freely along the 
copper-oxygen planes. Conversely, electrons placed on a complex
conducting lattice encount multiscale potential barriers, whose
height grows with absolute temperature $T$. (The effect of the 
barriers could be strengthened in the cuprates because of the 
electron absorbtion processes.) ``Loaded" with the positive mass 
(opposite to that of holes), electrons can enter into a superconducting 
state at the low enough temperatures $T \lesssim \hbar ^2 / 4m_e a^2$,
in accordance with condition (32). Support for this viewpoint can 
be found in the field-effect doping experiments with CaCuO$_2$ 
compound: \cite{Layer} While electron-type doping has shown 
the maximum value of $T_c$ around 34 K, hole-type doping
revealed an enhanced superconducting transition temperature $T_c$ 
near 89 K. Note that the field-effect geometry exploited in the
experiments in Ref. \cite{Layer} mimics the inherent microscopic 
organization of the copper oxides composed of the alternating 
conducting-insulating layers. 

We suggest that fracton coupled holes is the most 
probable ingredient of the high-temperature superconducting
phase in copper oxides. This proposal is somewhat different 
from the theoretical results of Buttner and Blumen, 
\cite{Buttner} who argued that such a phase might directly occur from 
the fracton-mediated electron-electron interactions. Nevertheless, 
the focus on fractons \cite{Buttner} as the effective glue that may 
confine the current-carrying particles in couples on a complex 
geometry is in accord with the basic ideas of our study.
Intrinsic multiscale inhomogeneities that might support the 
fracton modes in high-temperature superconductors have 
been studied experimentally in Ref. \cite{Mladen}

The direct observation of the fracton-hole interactions 
in the cuprate superconductors is an open challenge. Such interactions 
might be recognized from {\it meandering} of the charge stripes, 
a phenomenon addressed in Ref. \cite{Stripes} The meandering 
property has been strongly advocated in Ref. \cite{Evidence2}
from the novel in-plane transport anisotropy in copper oxides.  

\section{Ac Universality}

The complex percolating lattices underlying the fracton excitation 
spectrum (13) may leave an imprint on the ac conduction properties 
of the material above the superconducting transition temperature 
$T_c$. As is shown in Ref., \cite{PRB} self-organization of the 
medium to a state dominated by the critical percolation structures 
results in the universal behavior of ac conduction at the ``anomalous"
frequencies $\Omega$, for which the charge carriers walk on the
fractal. The defining feature of ac universality is independence
of the microscopic details of the pattern on which the random walk
takes place, and of the nature of the conducting mechanism operating
in the system (i.e., classical barrier crossing for ions and/or
quantum-mechanical tunneling for electrons). Ac universality can
be represented analytically by the power-law dependence of the
conductivity $\Sigma (\Omega)$ versus frequency $\Omega$:
\cite{Dyre}
\begin{equation}
\Sigma (\Omega)\, \propto\, \Omega ^{\eta}.
\eqnum{34}
\end{equation}
For the bulk conduction, the exponent $\eta$ lies typically between 
0.6 and 1.0. \cite{Dyre} Among other materials, the bulk ac universality
was recognized in ion and electron conducting polymers, amorphous and 
polycrystalline semiconductors, organic-inorganic composites, doped 
semiconductors at helium temperatures, etc. (see Refs. \cite{Dyre} 
and \cite{Rolla}, and references therein).

Ac universality (34) results from the self-similar behavior of 
the random walk processes on the fractal geometry. \cite{PRB} Given 
a path-connected fractal distribution with connectivity $\theta$,
one finds the mean-square displacement of the walker (e.g., of the 
charged particle placed on the disordered conducting array) after
time $t$ to be \cite{Gefen}
\begin{equation}
\left< \delta \chi ^2 (t) \right>\, \propto\, t^{2/(2 + \theta)}
~~~ (a\lesssim |\delta\chi| \lesssim\xi).
\eqnum{35}
\end{equation}
The probability to return to the starting point \cite{Procaccia}
scales with time as $p(t) \propto t^{- d_s / 2}$, where $d_s$ 
is the spectral dimension of the fractal, which simultaneously 
arises in the fracton DOS in Eq. (8). Near the threshold of percolation,
$d_s = {\cal {C}}$ and $p(t) \propto t^{- {\cal {C}} / 2}$.
In the limit of Euclidean geometry ($d_s \rightarrow d$, 
$\theta \rightarrow 0$), Eq. (35) recovers the conventional 
Einstein law $\left< \delta \chi ^2 (t) \right> \propto t$ for 
the particle diffusion. Generalized transport equations that
account for the ``strange" \cite{Nature} relation (35) have been 
considered in Refs. \cite{PRE01,Procaccia,Klafter,PhysicaD,Giona}

Based on the self-similar random walk model (35), 
Gefen {\it et al.} \cite{Gefen} demonstrated that ac 
conductivity of percolating fractal networks behaves as a 
power of frequency, $\Sigma (\Omega) \propto \Omega ^{\eta}$,
with the exponent $\eta = \mu / \nu (2 + \theta)$. This is in
accord with the dependence (34) customarily observed in real disordered
materials. Note that the quantity $\eta$ is a combination of the
indices $\mu$, $\nu$, and $\theta$, which mirror the percolative 
nature of the underlying fractal lattice. Making use of the 
expressions \cite{Gefen} $d_f = d - \beta / \nu$ and 
$\theta = (\mu - \beta) / \nu$, one finds
\begin{equation}
\eta\, =\, (\theta\, +\, d\, -\, d_f) / (2\, +\, \theta).
\eqnum{36}
\end{equation}
Equation (36) implies that the dependence $\Sigma (\Omega)$ is
controlled by the characteristic distance $|\delta\chi|$ traveled by 
a particle during the time $1/\Omega$. Considering expression (35),
one arrives at the scaling $\Omega\propto |\delta\chi| ^{- (2 + \theta)}$
which relates the applied frequency $\Omega$ to the displacement
$|\delta\chi|$.

For sufficiently low frequencies $\Omega \lesssim \xi ^{- (2 + \theta)}$,
the current-carrying particles cover distances $|\delta\chi| \gtrsim \xi$
during the period of $1/\Omega$. In this limit, the charge carriers feel
statistically homogeneous distribution of the conducting links,
corresponding to $d_f \rightarrow d$ and $\theta \rightarrow 0$.
From Eq. (36) one concludes that $\eta\rightarrow 0$ for 
$\Omega \lesssim \xi ^{- (2 + \theta)}$. Hence 
$\Sigma (\Omega) \propto \Omega ^{\eta}$ is almost 
frequency independent in the lower frequency range:
$\Sigma (\Omega) \rightarrow \Sigma (0) =$ const $(\Omega)$
for $\Omega \lesssim \xi ^{- (2 + \theta)}$, where $\Sigma (0)$
is dc conductivity.

For $\xi ^{- (2 + \theta)} \lesssim \Omega \lesssim a ^{- (2 + \theta)}$,
the charge transport processes are governed by the self-similar (fractal)
geometry of the underlying conducting set. In this regime, the value of 
$\eta$ obeys the topological constraints imposed by the percolation 
threshold character. In fact, combining Eqs. (12) and (36), we get 
($2\leq d\leq 5$):
\begin{equation}
\eta\, =\, (2\, -\, {\cal {C}}) /2\, +\, (d\, -\, 2) {\cal {C}} / 2 d_f.
\eqnum{37}
\end{equation}
For the plane lattices ($d=2$), the term depending 
on the Hausdorff dimension $d_f$ vanishes. Expression (37) is then 
reduced to the ``hyperuniversal" \cite{PRB} counterpart  
\begin{equation}
\eta\, =\, (2\, -\, {\cal {C}}) /2\, \approx\, 0.34\, \sim\, 1/3 
~~~ (d\, =\, 2),
\eqnum{38}
\end{equation}
involving solely the percolation constant ${\cal {C}}$. The exponent 
in Eq. (38) describes ac conduction of thin films of complex materials 
revealing a two-dimensional ($d=2$) self-similar organization. Ac 
conductivity of the self-similar (fractal) films scales with frequency 
$\xi ^{- (2 + \theta)} \lesssim \Omega \lesssim a ^{- (2 + \theta)}$
as $\Sigma (\Omega) \propto \Omega ^{1/3}$, where we neglected the small
deviation of the percolation constant ${\cal {C}} \approx 1.327$
from the mean-field value 4/3. 

The ``hyperuniversal" \cite{PRB} behavior
$\Sigma (\Omega) \propto \Omega ^{1/3}$ in two dimensions should 
be considered as a specific prediction of our model: To our best 
knowledge, reliable experimental data on ac conduction of the fractal
films are almost absent in the literature. Direct observational
verification of the scaling $\Sigma (\Omega) \propto \Omega ^{1/3}$ 
might be performed, e.g., through a charge injection in the 
field-effect-transistor geometry using the pioneering methods
discussed in Refs. \cite{Molecular,Schon,Layer} 

In higher dimensions ($3\leq d\leq 5$), the ``Hausdorff" 
term $(d - 2) {\cal {C}} / 2 d_f$ on the right of expression (37) 
apparently contributes to the exponent $\eta$. Considering the constraint 
(17) on the parameter $d_f$, we locate the value of $\eta$ in the interval
$\eta _{\min}\leq\eta\leq\eta _{\max}$, with 
\begin{equation}
\eta _{\min}\, =\, 
(2\, -\, {\cal {C}}) /2\, +\, (d\, -\, 2) {\cal {C}} / 2 {\cal {S}} _d,
\eqnum{39}
\end{equation}
\begin{equation}
\eta _{\max}\, =\, 
1\, +\, (d\, -\, 3) {\cal {C}} / 2.
\eqnum{40}
\end{equation}
Here, ${\cal {S}} _d = \ln (3^d - 1) / \ln 3$ 
is the Hausdorff dimension of the Cantor cheese in $E^d$. 
In three dimensions ($d=3$), one finds $\eta _{\min}\leq\eta\leq 1$, 
where $\eta _{\min} = (2 - {\cal {C}}) /2 + {\cal {C}} / 2 {\cal {S}} _3$.
Numerically, $\eta _{\min} \approx 0.56$, in close agreement with the 
observational result $0.6\lesssim \eta\leq 1.0$. \cite{Dyre} 

In view of inequality (2), from Eq. (37) we also get
\begin{equation}
\partial \eta / \partial T\, =\, 0
~~~ (d\, =\, 2),
\eqnum{41}
\end{equation} 
\begin{equation}
\partial \eta / \partial T\, \leq\, 0
~~~ (d\, =\, 3).
\eqnum{42}
\end{equation}
Hence, in three dimensions, $\eta$ generally
increases with decreasing temperature $T$, a phenomenon
often observed in real disordered materials. \cite{Dyre} 

We emphasize that ac universality (34), although effective 
{\it above} the typical superconducting transition temperature 
$T_c$, is an integral part of the fracton scenario as it derives 
from the same general principles basing on the percolative geometry
of the underlying conducting lattice. Ac universality might serve as 
an indication that the material posseses the complex enough microscopic 
organization that supports a wide fracton spectrum. Ac-universal 
materials could thereby be plausible candidates into the strange 
superconductors dominated by the fracton-mediated interactions 
between the current-carrying particles. A test on ac universality 
might help verify the fracton nature of the pairing mechanism 
potentially operating in organic polymers and copper oxides.

\section{Summary and Conclusions}

In spite of the diverse superconducting transition temperatures,
self-assembling organic polymers and copper-oxide compounds bear
important features in common. In fact, the polymers superconduct in 
the field-effect-transistor geometry at high gate-induced charge-carrier
concentrations. The superconducting state is supported by self-organization 
mechanisms governing the kinetics of the polymer system. Owing to 
self-organization, the polymers form thin regular films, on which 
the superconducting electron fluid condensates. 

The cuprate superconductors, in their turn, 
operate like multi-layer transistors due to the intrinsic field-effect 
crystalline configuration (i.e., conducting copper-oxygen planes confined 
between charge-absorbing insulating substrates). Self-organization processes 
appear in the charge-stripe order enabling the current-carrying 
particles flow along the copper-oxygen planes without resistance. 

Both polymers and copper oxides are ``strange" superconductors
as they involve unconventional (other than BCS) pairing processes
underlying the superconducting phase. In either of the materials, 
the (quasi)particles mediating the interaction between the charge
carriers are suggested to be fractons, quantum oscillations of 
fractal lattices. 

The interaction processes involving fractons allow for larger
energies than that of the phonons, leading to a possibility of
the high-temperature superconducting phases in complex materials. 
For the cuprate compounds, we estimate the typical superconducting 
transition temperature to be around $T_c\sim 150$ K, in good agreement 
with the experimental values. We imply that the marginal ingredient 
of the corresponding high-temperature superconducting phase are 
fracton coupled holes that condensate in the conducting copper-oxygen 
planes owing to the specific field-effect-transistor geometry of the 
cuprate superconductors. For the gate-induced superconductivity in the 
self-assembling polymers, we find much smaller transition temperatures 
of $\sim (2-3)$ K, provided the charge carriers are fracton 
coupled electrons, not holes. Such a modest value is due to the 
limitations imposed by the electron tunneling processes on the 
braided polymer structures. We speculate that replacing electrons 
with holes helps circumvent the tunneling restrictions. The fracton 
model then promises enhanced onset temperatures for the hole-doped systems 
when compared to their electron-doped counterparts. The hint is that 
hole-doped complex materials whose structural characteristics meet 
a wide fracton spectrum, might be feasible candidates in the
high-temperature superconductors. Support for this suggestion 
partially comes from the field-effect doping experiments 
discussed in Ref. \cite{Layer} Hole doping might thereby open 
a perspective on the superconducting systems having large
transition temperatures $T_c \gtrsim 150$ K.

We found that the fracton wave function contains a power-law 
tail stretching far beyond the central core region. The power-law 
constituent supports interactions between the charge carriers 
throughout the self-similar (fractal) domain of the conducting system. 
In the harmonic approximation, the fracton wave function derives from a 
generalized kinetic equation with the integer time and fractional 
real-space derivatives. The latter is given by the Riesz operator 
addressing the fractional generalization of the Laplacian. The
introduction of the fractional kinetic equation for fracton
excitations follows the paradigm of the {\it strange dynamics}
\cite{Nature}, one of the biggest open challenges in the modern
nonlinear physics. The parameters of the fracton wave function
describing the shape of the core and tail regions could be obtained
from the unconventional, topological analysis basing on the concept 
of the {\it fractional manifold}. \cite{PRE97} The fractional kinetics 
approach supplied with the topological guide to fractal objects 
yields insight into the fundamental dynamical properties 
underlying fracton statistics. 

The formation of self-similar (fractal) lattices associated with
a well-defined branch of fracton excitations affects ac conduction 
properties of the material in the classical energy range (i.e., above 
the typical superconducting transition temperature $T_c$). Based on the
topological arguments, we demonstrated the universal character of
ac conduction in fractal geometries, which does not depend on the 
details of the lattice, nor on the nature of the microscopic conducting 
mechanism. This universal character is contained in the scaling 
$\Sigma (\Omega) \propto \Omega ^{\eta}$, with the exponent $\eta$ 
related to the fundamental topological parameters characterizing the 
fractal distribution, such as the percolation constant \cite{PRE97} 
and the Hausdorff dimension of the Cantor cheese. \cite{Schroeder} 
In three dimensions (i.e., for the bulk conduction), the value of 
$\eta$ varies from approximately 0.56 to 1.0, in precise agreement 
with the experimental range $0.6\lesssim \eta\leq 1.0$. \cite{Dyre} 
In two dimensions (i.e., for the plane fractal lattices), 
a ``hyperuniversal" \cite{PRB} value $\eta\approx 0.34\sim 1/3$ 
is recognized, leading to the scaling $\Sigma (\Omega) \propto 
\Omega ^{1/3}$ in a wide temperature range $T\gtrsim T_c$. This 
behavior should be considered as a specific theoretical prediction 
of our model. Observational verification of the scaling $\Sigma (\Omega) 
\propto \Omega ^{1/3}$ in plane fractal geometries might shed 
light on the electric properties of the complex materials. 

It might be extremely interesting to verify the scaling 
$\Sigma (\Omega) \propto \Omega ^{1/3}$ for cuprate oxides 
and self-assembling organic polymers, whose microscopic organization
seems to support plane fractal lattices. An observation of the scaling
$\Sigma (\Omega) \propto \Omega ^{1/3}$ might be a strong argument in 
favour of the fracton-mediated pairing processes operating in the 
strange superconductors. Extensive studies in this direction might 
be strongly advocated.

%%%%%%%%%%%%%%%%%%%%%%%%%%%%%%%%%%%%%%%%%%%%%%%%%%%%%%%%%%%%%%%%%%%%%%%%%%%%%%
%%%%%%%%%%%%%%%%%%%%%%%%%%%%%%%%%%%%%%%%%%%%%%%%%%%%%%%%%%%%%%%%%%%%%%%%%%%%%%

%%% \acknowledgements
We are grateful to the anonymous referee for the careful 
reading of this paper and for the stimulating suggestions.
This study was carried out when A. V. M. stayed at Risoe on a Grant from the
Graduate School in Nonlinear Science. Partial support was received from
Danish Natural Science Foundation (SNF), INTAS Project 97-1612, and 
RFFI Grants 00-02-17127, 00-15-96631, and 02-02-06541.
%%%%%%%%%%%%%%%%%%%%%%%%%%%%%%%%%%%%%%%%%%%%%%%%%%%%%%%%%%%%%%%%%%%%%%%%%%%%%%
%%%%%%%%%%%%%%%%%%%%%%%%%%%%%%%%%%%%%%%%%%%%%%%%%%%%%%%%%%%%%%%%%%%%%%%%%%%%%%
%%%%%%%%%%%%%%%%%%%%%%%%%%%%%%%%%%%%%%%%%%%%%%%%%%%%%%%%%%%%%%%%%%%%%%%%%%%%%%

%%%%%%%%%%%%%%%%%%%%%%%%%%%%%%%%%%%%%%%%%%%%%%%%%%%%%%%%%%%%%%%%%%%%%%%%%%%%%%
%%%%%%%%%%%%%%%%%%%%%%%%%%%%%%%%%%%%%%%%%%%%%%%%%%%%%%%%%%%%%%%%%%%%%%%%%%%%%%

\end{document}